\documentclass{optica-article}

\journal{opticajournal} 

\articletype{Research Article}

\usepackage{lineno}
\usepackage{physics}

\begin{document}

\title{Generation of multi-photon Fock states at telecommunication wavelength using picosecond pulsed light}

\author{Tatsuki Sonoyama,\authormark{1} Kazuma Takahashi,\authormark{1} Tomoki Sano, \authormark{1} Takumi Suzuki, \authormark{1} Takefumi Nomura,\authormark{1} Masahiro Yabuno,\authormark{2} Shigehito Miki,\authormark{2} Hirotaka Terai,\authormark{2} Kan Takase,\authormark{1,3} \\ Warit Asavanant,\authormark{1,3} Mamoru Endo,\authormark{1,3} and Akira Furusawa\authormark{1,3}}

\address{\authormark{1} Department of Applied Physics, School of Engineering, The University of Tokyo,\\
7-3-1 Hongo, Bunkyo-ku, Tokyo 113-8656, Japan\\
\authormark{2} Advanced ICT Research Institute, National Institute of Information and Communications Technology, \\
588-2 Iwaoka, Nishi-ku, Kobe, Hyogo 651-2492, Japan\\
\authormark{3}Optical Quantum Computing Research Team, RIKEN Center for Quantum Computing,\\
2-1 Hirosawa, Wako, Saitama 351-0198, Japan}

\email{\authormark{*}sonoyama@alice.t.u-tokyo.ac.jp}


\begin{abstract*} 
Multi-photon Fock states have diverse applications such as optical quantum information processing. For the implementation of quantum information processing, it is desirable that Fock states be generated within the telecommunication wavelength band, particularly in the C-band (1530-1565 nm). This is because mature optical communication technologies can be leveraged for the transmission, manipulation, and detection. Additionally, to achieve high-speed quantum information processing, it is desirable for Fock states to be generated in short optical pulses, as this allows embedding lots of information in the time domain. In this paper, we report the first generation of picosecond pulsed multi-photon Fock states (single-photon and two-photon states) in the C-band with Wigner negativities, which are verified by pulsed homodyne tomography. In our experimental setup, we utilize a single-pixel superconducting nanostrip photon-number-resolving detector (SNSPD), which is expected to facilitate the high-rate generation of various quantum states. This capability stems from the high temporal resolution of SNSPDs (50 ps in our case) allowing us to increase the repetition frequency of pulsed light from the conventional MHz range to the GHz range, although in this experiment the repetition frequency is limited to 10 MHz due to the bandwidth of the homodyne detector. Consequently, our experimental setup is anticipated to serve as a prototype of a high-speed optical quantum state generator for ultrafast quantum information processing at telecommunication wavelength.
\end{abstract*}

\section{Introduction}
Fock states $\ket{n}$, also known as photon number states, are quantum states defined as eigenstates of the photon number operator $\hat{n}=\hat{a}^\dagger\hat{a}$, representing quantum states with a definite number of photons excited in the electromagnetic field. Beyond their intrinsic physical interest, Fock states hold diverse applications in quantum metrology, quantum communication, and quantum computation \cite{PhysRevLett.71.1355,Couteau2023,Knill2001}. Over the past decades, experimental techniques for generating Fock states have been realized through various methods \cite{Arakawa-quantumdot, Nogues1999, Aharonovich2016}. Among all, the approach called heralding scheme utilizing quantum entanglement and photon-number-resolving (PNR) detection has advantages in that it enables high-purity and multi-photon Fock state generation in a well-defined spatial and temporal mode \cite{PhysRevLett.96.213601, Cooper:13, Bouillard:19,Kawasaki:22}.

Considering applications in optical quantum information processing, it is desirable to generate Fock states in the C-band wavelength where losses in fiber-based systems are low and the utilization of optical communication technologies is anticipated \cite{Inoue-5G} for its transmission, manipulation and detection. Furthermore, in order to realize ultrafast optical quantum information processing, it is preferable for optical quantum states to be generated in as short an optical wave packets as possible. When using continuous wave (CW) light in heralding scheme, the generated state's optical wave packet is typically on the order of nanoseconds \cite{PhysRevResearch.5.033156, Yukawa:13}, whereas employing pulsed light enables the generated state's optical pulse to be on the order of picoseconds or even shorter. 
Previous studies report the generation of Fock states on subpicosecond optical wave packet using pulsed light source at near-infrared wavelength with Wigner negativity \cite{PhysRevLett.96.213601, Cooper:13, Bouillard:19}. On the other hand, examples of pulsed Fock state generation within the telecommunication wavelength band are scarce. While there are reports on g(2) measurements for single photons \cite{Bock:16}, there are still no reports on the generation of pulsed Fock states at telecommunication wavelength that maintain Wigner negativity without any loss correction, which is crucial for quantum information processing \cite{PhysRevLett.109.230503}.

In this study, we generate picosecond pulsed single-photon and two-photon states at around 1545.32 nm (C-band) using a type-I\hspace{-1.2pt}I periodically poled LiNbO$_{3}$ (PPLN) waveguide and a superconducting nanostrip photon-number-resolving detector (SNSPD). Subsequently, we reconstruct the Wigner functions of the generated states through quantum state tomography using homodyne detection and confirm the Wigner negativity without any loss correction as an indicator of non-classicality. 

The experimental setup holds promise as a prototype for a high-speed quantum state generator in optical quantum information processing for the following two reasons. First, it can be applied not only to the generation of Fock states but also to various non-Gaussian state generations as shown in Fig.1. This versatility arises from utilizing Einstein–Podolsky–Rosen (EPR) states generated by a type-I\hspace{-1.2pt}I PPLN waveguide. For instance, by applying displacement operations to the photon detection mode of the EPR state, cubic phase states \cite{PhysRevA.64.012310} can be generated. Additionally, by inserting a half-wave plate (HWP) in front of the polarizing beam splitter (PBS) that separates the EPR state spatially, Schr\"{o}dinger cat states \cite{PhysRevLett.120.073603} can be generated. This is a decisive difference from a similar method called photon subtraction using squeezed light sources \cite{Endo:23}. Secondly, the use of single-pixel SNSPD \cite{Cahall:17,Endo:21} with low timing jitter allows for further acceleration of state generation rates. In many previous studies, pseudo-photon-number-resolving measurements have been employed by parallelizing single-photon detectors for Fock-state generation \cite{PhysRevLett.96.213601, Cooper:13, Bouillard:19}. However, this approach inevitably leads to errors when multiple photons enter the same single-photon detector \cite{Provaznik:20}. Because it necessitates operation in the weak pumping regime where the squeezing level of the EPR state is low enough so that the error probability is kept low. In contrast, the single-pixel PNR detector used in this study does not suffer from such errors, allowing operation even under strong pumping conditions and thus enabling faster state generation rates. Furthermore, compared to conventional single-pixel PNR detectors such as the Transition Edge Sensor (TES) \cite{TES-ns}, the SNSPD used in this study has a time resolution that is approximately 100 times faster, on the order of a few tens of picoseconds \cite{SNSPD}. Consequently, the repetition frequency of pulsed light source can be increased accordingly. In this experiment, the bandwidth of the homodyne detector is around several tens of MHz, which prevents us from increasing the repetition frequency. However, by utilizing recently developed high-speed homodyne detectors with bandwidths of several tens of GHz \cite{Inoue-5G,kawasaki2024highrate}, it is theoretically possible to increase the repetition frequency up to several tens of GHz. From these, our experimental setup has the potential for the high-rate generation of various optical quantum states at telecommunication wavelength and in the picosecond optical wavepacket. Thus, this study serves as a demonstration experiment for a prototype of quantum state generator used in ultrafast optical quantum information processing.

\section{Experiment}
\subsection{Fock state generation by a heralding scheme}
\begin{figure}[htbp]
  \centering
  \includegraphics[width=\textwidth]{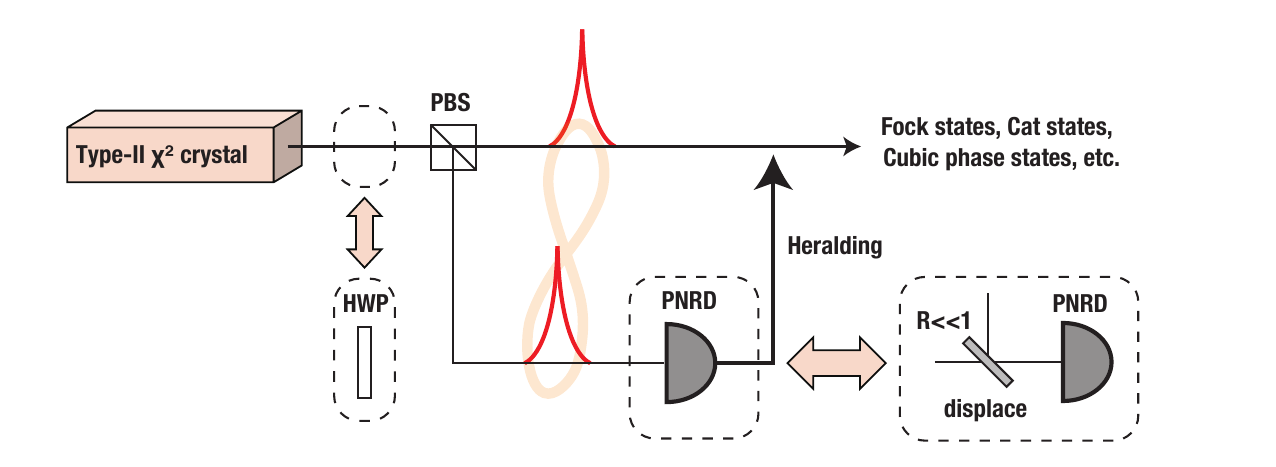}
\caption{The schematic diagram of the optical quantum states generation using the heralding method. In the simplest example, Fock states are generated by performing photon-number-resolving detection (PNRD) on one mode of the EPR states. Various non-Gaussian states, such as Schr\"{o}dinger's cat states or cubic phase states can be also generated by inserting a half-wave plate (HWP) before the polarizing beamsplitter (PBS) or by applying displacement operations before the PNRD.}
\end{figure}

Spontaneous parametric down-conversion (SPDC) is a second-order nonlinear optical process, whose Hamiltonian is given by $\hat{\mathcal{H}} = \chi (\hat{a}^{\dagger}_{\rm i} \hat{a}^{\dagger}_{\rm s} \beta_{\rm p} + \hat{a}_{\rm i} \hat{a}_{\rm s} \beta^{*}_{\rm p})$. Here, $\hat{a}_{\rm i}$, $\hat{a}_{\rm s}$ represent the annihilation operators for idler and signal light and $\beta_{\rm p}$ represent the amplitude of pump light, which is sufficiently strong. In the case where the input idler and signal are in a vacuum state, the output state $\ket{\psi}$ of this process is expressed as follows, 
\begin{equation}
    \ket{\psi} = \exp(-i\frac{\hat{\mathcal{H}}}{\hslash}t)\ket{0}_{\rm i}\ket{0}_{\rm s} = \frac{1}{\cosh{r}} \sum_{n=0}^{\infty} \tanh^{n}{r} \ket{n}_{\rm i}\ket{n}_{\rm s},
\end{equation}

where $r$ is the squeezing parameter proportional to the coefficient $\chi$ and the amplitude of the pump light $\beta_{\rm p}$. The output state $\ket{\psi}$ is referred to as EPR state or two-mode squeezed state, where the signal and idler exhibit quantum entanglement. This process can be understood as the conversion of pump photons into idler and signal photons, satisfying the energy conservation $\nu_{p} = \nu_{s} + \nu_{i}$, where $\nu$ is the frequency of light. In the type-I\hspace{-1.2pt}I nonlinear optical crystal used in this experiment, signal and idler light are generated in different polarization modes, allowing spatial separation of the two modes using a polarizing beamsplitter (PBS). When we detect $n$ photons on the idler mode, Fock state $\ket{n}$ is generated on the signal mode with the probability $p(n) = \tanh^{2n}{r}/\cosh^2{r}$. Furthermore, by performing displacement operations on the idler or by inserting half-wave plate before the PBS, various quantum states such as cubic phase states or Shr\"{o}dinger cat states can be generated \cite{PhysRevA.64.012310,PhysRevLett.120.073603} as shown in Fig.1. 

Experimentally, optical losses in the signal and the idler both deteriorate the generated states in different ways. The losses in the idler have the effect of increasing the average photon number of the signal, leading to a classical mixture of states with photon numbers of $n$ and greater than $n$. This occurs because an idler with $n+k$ photons can sometimes be recognized as $n$ photons by PNRD. Additionally, the generation rate decreases. On the other hand, the losses in the signal have the effect of decreasing the average photon number, resulting in a classical mixture of states with photon numbers of $n$ and less than $n$.

\subsection{Experimental setup}
\begin{figure}[htbp]
  \centering
  \includegraphics[width=\textwidth]{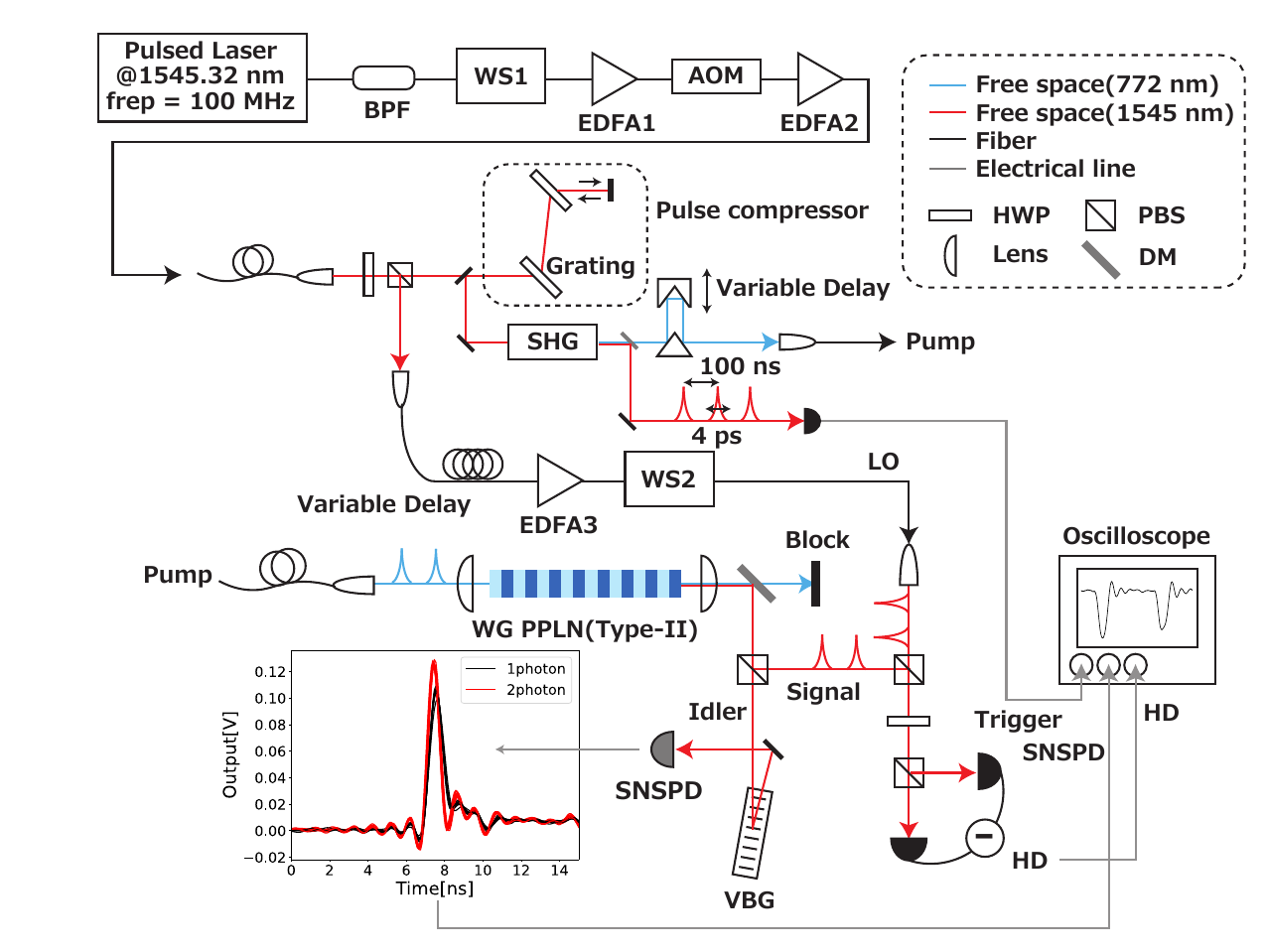}
\caption{The detailed experimental setup for the generation of Fock states. The graph shows the overwritten output of the SNSPD’s signals. The red lines correspond to the two-photon detection event and the black lines correspond to the single-photon detection event. BPF: Bandpass Filter, WS: Waveshaper, AOM: Acousto-Optic Modulator, EDFA: Erbium-Doped Fiber Amplifier, HWP: Half-wave plate, PBS: Polarizing Beamsplitter, DM: Dichroic Mirror, SHG: Second Harmonic Generation, LO: Local Oscillator, WG: Waveguide, VBG: Volume Bragg Grating, SNSPD: Superconducting Nanostrip Photon-number-resolving Detector, HD: Homodyne Detector.}
\end{figure}

Fig.2 shows the experimental setup. Initially, a pulsed laser with a repetition frequency of 100 MHz and a wavelength of 1545.32 nm is used as a light source. To achieve a desirable pulse width of a few picoseconds in this experiment, a bandpass filter (BPF) 2 nm bandwidth is introduced. Subsequently, the waveform is further shaped and chirped with a waveshaper (WS) to suppress the peak power of the pulsed light. After passing through an Erbium-Doped Fiber Amplifier (EDFA), the repetition frequency of pulsed light is reduced from 100 MHz to 10 MHz using a pulse picker based on an AOM (Acousto-Optic Modulator). After that, the light is split into two using a HWP and a PBS. One port serves as the local oscillator (LO) for homodyne detection (HD), while the other is designated as the pump light for SPDC. The path for LO involves variable delayline to match the time delay between the LO light and the signal light. In the pulsed HD, where optimization of the temporal waveform of LO is crucial, the waveform is shaped with a waveshaper. On the other path, after passing through a dispersion compensator using a pair of gratings, the pulsed light enters a fan-out periodically-polled MgO-doped near-stoichiometric LiTaO3 crystal for second harmonic generation (SHG), resulting in the generation of pump light at the wavelength of 772.66 nm. The duration of the pulsed light source before the SHG crystal is measured to be 4 ps. After a variable delayline, the light is input into the type-I\hspace{-1.2pt}I waveguide (length is $= 45$ mm). The average power of input pump light can be tuned from 0 {\textmu}W to 315 {\textmu}W, which corresponds to the pulse energy of $31.5$ pJ and the peak power of $11.7$ W. The EPR states generated by SPDC are spatially separated by a PBS. One mode (idler) passes through a volume Bragg grating (VBG) for frequency filtering of 0.2 nm bandwidth and is detected by the PNR detector SNSPD. The setup of the SNSPD is mostly the same as \cite{Endo:21}, except for the wavelength and the electrical high-pass filter (cutoff frequency $f_c = 3.8$ GHz) introduced in the output circuit to distinguish the rise time of the output signal. The other mode (signal) is incident into the HD system. Using a PBS, a HWP and another PBS, the LO and the signal are combined at a 50:50 ratio and measured by a homodyne detector. Here, the measured temporal mode in the pulsed HD is determined by the temporal mode of LO, which is optimized by the waveshaper and the variable delayline. The duration of the optimized LO is measured to be around 13 ps, which also corresponds to the duration of the generated states. The reason of the difference between the duration of input pulsed light and that of the generated state is that the temporal mode of the generated state is determined not only by the spectrum of the pump but also by the phase-matching bandwidth of the crystal and the frequency filter of the idler. Pulsed HD are conducted 10,000 times for both single-photon and two-photon states, and the probability distribution of quadratures is measured. The count rates for single-photon in this experiment are 4,100 cps (pump power = 75{\textmu}W), 8,200 cps (150 {\textmu}W), 10,500 cps (225 {\textmu}W), 12,000 cps (315 {\textmu}W) and the count rates for two-photon is 50 cps (300 {\textmu}W). 

Fig. 2 shows how the quadrature is extracted from the HD signal. The output signal of the HD is input to the oscilloscope together with the output signal of the SNSPD. Here, these signals are triggered by an output signal of the photodetector which measures the classical pulsed light. Fig. 2 (a) shows an actual SNSPD signal detecting two photons. In this setup, voltage thresholds are set to distinguish between zero-photon detection, single-photon detection and two-photon detection. For the generation of single-photon states, only HD signals exceeding the threshold for single-photon detection are extracted. Similarly, for the generation of two-photon states, only HD signals exceeding the threshold for two-photon detection are extracted. In Fig.3 (b), the HD signals for a two-photon detection event is overwritten 20 times. The presence of an offset in the HD signals is likely due to factors such as the unbalance of the beam splitter and differences in the responses of the left and right photodetectors. We have confirmed that the error of branching ratio is within 0.5\% when compared to cases with a similar offset on the positive side, thus we believe that this influence can be sufficiently ignored. The variance of the HD signals is shown in Fig.3 (c). One pulse have a larger variance than the adjacent pulses, which corresponds to the two-photon states. In this experiment, the voltage values at the time indicated by the vertical dotted lines are extracted and normalized using shot noise to represent the quadratures.

\begin{figure}[htbp]
  \centering
  \includegraphics[width=0.75\textwidth]{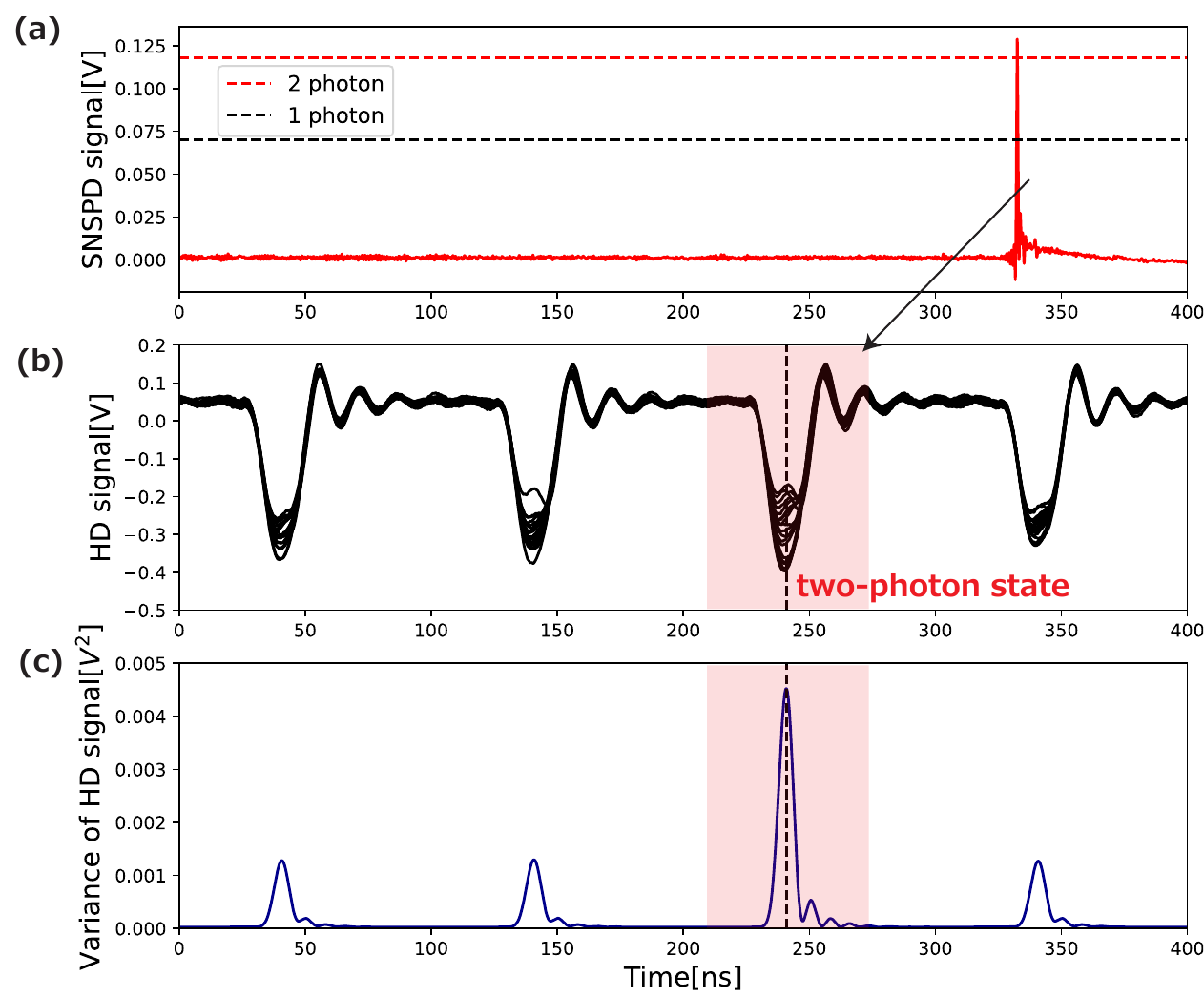}
\caption{(a) One of the SNSPD's signals detecting two photons. The black dashed line represents the threshold for detecting one photon, and the red dashed line represents the threshold for detecting two photons. (b) The HD signals when the SNSPD detects two photons. This figure overwrites the signals 20 times. The quadrature is measured every 100 ns, and the third pulse corresponds to the two-photon detection event in this case. Here, the SNSPD signal and the HD signal are triggered by the electrical signal of the photodetector measuring the pulsed light source. (c) The variance of the homodyne measurement values, which is calculated from all HD signals of the two-photon detection events. It can be observed that the variance is significantly larger only for the pulse corresponding to two-photon detection events.}
\end{figure}

\section{Analysis and Results}
We conduct homodyne measurements 10,000 times each for the output states when a single photon and when two photons are detected by the SNSPD, respectively. In the case of two-photon detection, occasionally, offset signals are superimposed due to the SNSPD detecting a single photon in the preceding one or two pulses, causing the originally single-photon detection to be misidentified as two-photon detection. To prevent such false detections, post-processing is performed to remove events where photon detection occurred in one or two pulses preceding the actual two-photon detection event. Among the 10,000 two-photon detection events, such events occurred 498 times. After these post-processing steps, we use maximum likelihood estimation \cite{ILvovsky_2004} to reconstruct the density matrices of the generated quantum states from 10,000 sets of quadrature data for single-photon detection and 9,502 sets of quadrature data for two-photon detection. Since the Fock states generated in this study are phase-insensitive states, it is assumed that the estimation of the density matrix is also phase-insensitive. 

Fig.4 shows the measurement results for the vacuum state measured by blocking the signal, the single-photon state obtained with a pump light of 75 {\textmu}W, and the two-photon state obtained with a pump light of 315 {\textmu}W. Fig.4 (a-c) show the probability distributions of the quadratures obtained from the experiment and those obtained from the reconstructed density matrices. The variances of the quadrature of single-photon and two-photon states are larger compared to the vacuum states, and characteristic structures of Fock states' wavefunctions can be observed. Moreover, the probability distribution of the quadrature obtained from the reconstructed density matrix is in good agreement with the experimental values, indicating successful estimation of the density matrices. Fig.4 (d-e) and Table.1 show the diagonal components of the reconstructed density matrices, namely the photon number distributions.  In the quantum states obtained from single-photon detection, the single-photon component is the largest at 62\%, while in the quantum states obtained from two-photon detection, the two-photon component is the largest at 41\%. Photon number components fewer than the detected photon number are mainly caused by losses on the signal side, while those greater than the detected photon number are mainly caused by losses on the idler side. Additionally, since we set the two-photon detection threshold voltage as high as possible to avoid mixing single-photon detection events in two-photon detection events, there is a possibility that we are discarding two-photon detection events that should have been detected, leading to an increase in the proportion of higher photon number detection events. This could be one of the reasons of higher photon number components in the generated two-photon states. Fig.4 (f-g) show the Wigner functions of single-photon states and two-photon states. The Wigner functions respectively have negative values, which are indicators of non-classicality: $W (0,0) = -0.081 \pm 0.007$ for single-photon states and $W (0,0.65) = -0.0082 \pm 0.0034$ for two-photon states. No loss corrections are made, and the error bars are estimated using the bootstrap method. For the single-photon state, although only the case with a pump light of 75 {\textmu}W is illustrated in Fig.4, negative values of the Wigner function have been also confirmed for intensities of 150 {\textmu}W, 225 {\textmu}W, and 315 {\textmu}W as well. From these results, we conclude that we succeed in the generation of multi-photon Fock states.

\begin{figure}[htbp]
  \centering
  \includegraphics[width=\textwidth]{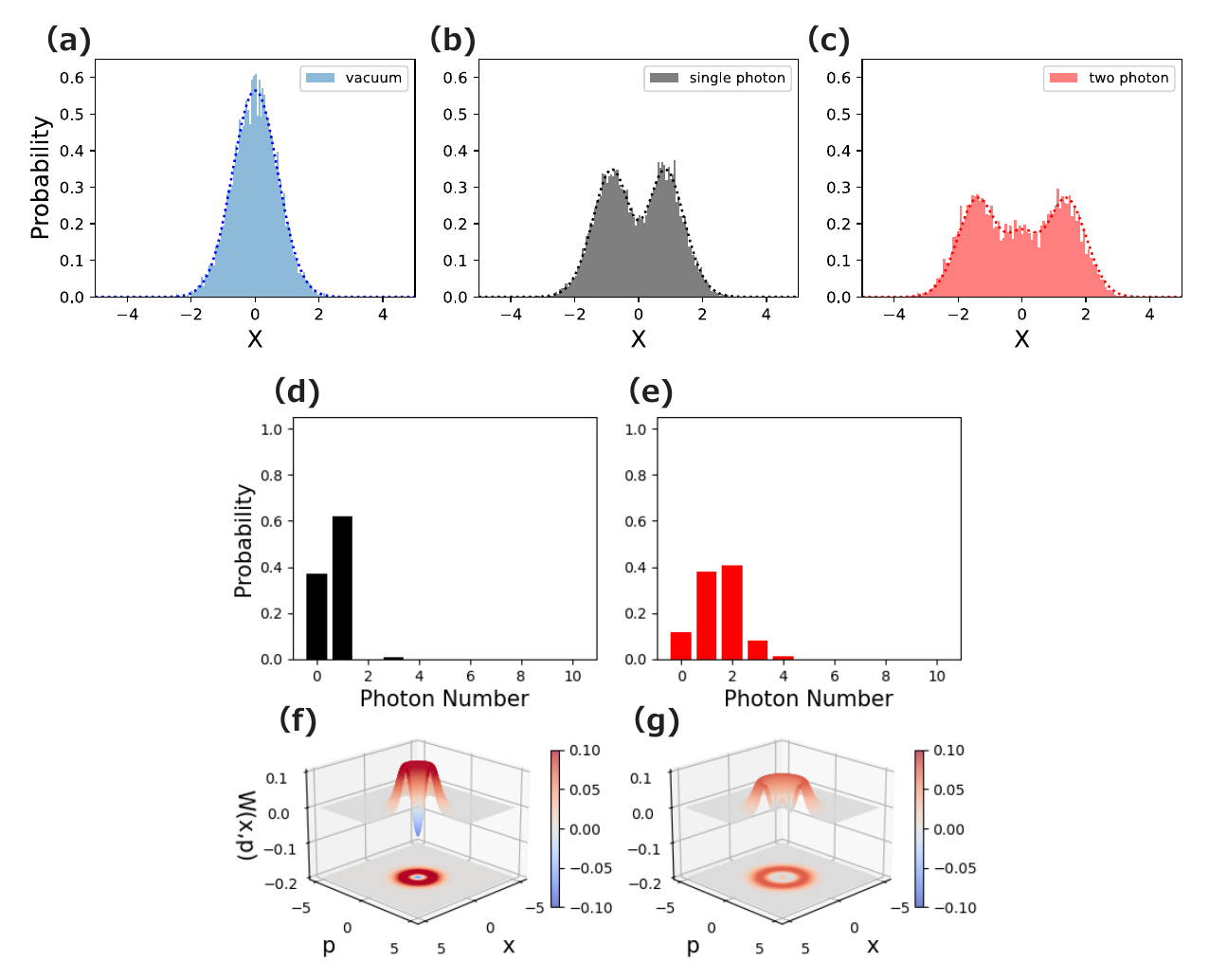}
\caption{The results of measurement of vacuum states, single-photon states (pump power = 75 {\textmu}W), two-photon states (pump power = 300 {\textmu}W). (a-c) The probability distribution of quadratures obtained from the experiment (histogram) and that obtained from the estimated density matrices (dashed line) for (a) vacuum states, (b) single-photon states, (c) two-photon states. For the calculation of theoretical probability distribution (dashed line) of vacuum states in (a), we use $\dyad{0}$ as the density matrix. (d-e) The estimated photon-number distributions without any loss corrections for (d) single-photon states and (e) two-photon states. (f-g) The estimated Wigner functions without any loss corrections for (f) single-photon states and (g) two-photon states. The estimated Wigner functions have negative values, $W(0,0)= -0.081 \pm 0.007$ for single-photon states, and $W(0,0.065)=-0.082 \pm 0.0034$ for two-photon states.}
\end{figure}

\section{Conclusion}
We have successfully generated single-photon and two-photon states in the telecommunication wavelength band using picosecond optical pulses. Furthermore, we have successfully observed negative values of the Wigner function without any loss correction through pulse homodyne tomography. Our experimental setup utilizes a type-I\hspace{-1.2pt}I PPLN waveguide to generate EPR states and a recently developed high-speed PNR detector, the SNSPD. This setup enables high-speed generation of various optical non-Gaussian quantum states, making it directly applicable to ultrafast optical quantum computation. We believe that our experiment serves as a prototype for a high-speed optical quantum state generator in ultrafast optical quantum computers.

\begin{table}[htbp]
\begin{tabular}{c|ccccc}
    \hline
     & $P(0)$ & $P(1)$ & $P(2)$ & $P(3)$ & $P(4)$ \\
    \hline
    Single-photon state& 37.2$\pm 0.8$ \% & 62.0$\pm 1.3$ \% & 0.0$\pm 1.2$ \% & 0.8$\pm 0.6$ \% & 0.0$\pm 0.0$ \%\\
    Two-photon state& 11.9$\pm 0.8$ \% & 38.2$\pm 1.1$ \% & 40.8$\pm 1.2$ \% & 7.9$\pm 1.2$ \% & 1.3$\pm 0.7$ \%\\
    \hline
   \end{tabular}
\centering \caption{Estimated photon-number distribution of the generated single-photon state and the generated two-photon state. $P(i)$ corresponds to the probability of $i$ photons. The standard deviation of the error is calculated by the bootstrap method.}
\end{table}

\section*{Funding}
Japan Science and Technology Agency (JPMJMS2064, JPMJPR2254); Japan Society for the Promotion of Science (18H05207, 20K15187, 22K20351, 23KJ0518, 23K13038)

\section*{Acknowledgments}
T.Sonoyama, K.Takahashi, T.N, and T.Suzuki acknowledge financial supports from The Forefront Physics and Mathematics Program to Drive Transformation (FoPM). M.E., K. Takase, and W.A. are supported by the Research Foundation for Opto-Science and Technology. The authors acknowledge UTokyo Foundation and donations from Nichia Corporation. 

\section*{Disclosures}
The authors declare no competing financial interests.

\section*{Data availability}
The raw data are not publicly available at this time but may be obtained from the authors upon reasonable request.

\bibliography{citation}

\begin{thebibliography}{10}
\newcommand{\enquote}[1]{``#1''}

\bibitem{PhysRevLett.71.1355}
M.~J. Holland and K.~Burnett, \enquote{Interferometric detection of optical phase shifts at the heisenberg limit,} {\protect\JournalTitle{Phys. Rev. Lett.}} \textbf{71}, 1355--1358 (1993).

\bibitem{Couteau2023}
C.~Couteau, S.~Barz, T.~Durt, \emph{et~al.}, \enquote{Applications of single photons to quantum communication and computing,} {\protect\JournalTitle{Nature Reviews Physics}} \textbf{5}, 326--338 (2023).

\bibitem{Knill2001}
E.~Knill, R.~Laflamme, and G.~J. Milburn, \enquote{A scheme for efficient quantum computation with linear optics,} {\protect\JournalTitle{Nature}} \textbf{409}, 46--52 (2001).

\bibitem{Arakawa-quantumdot}
Y.~Arakawa and M.~J. Holmes, \enquote{{Progress in quantum-dot single photon sources for quantum information technologies: A broad spectrum overview},} {\protect\JournalTitle{Applied Physics Reviews}} \textbf{7}, 021309 (2020).

\bibitem{Nogues1999}
G.~Nogues, A.~Rauschenbeutel, S.~Osnaghi, \emph{et~al.}, \enquote{Seeing a single photon without destroying it,} {\protect\JournalTitle{Nature}} \textbf{400}, 239--242 (1999).

\bibitem{Aharonovich2016}
I.~Aharonovich, D.~Englund, and M.~Toth, \enquote{Solid-state single-photon emitters,} {\protect\JournalTitle{Nature Photonics}} \textbf{10}, 631--641 (2016).

\bibitem{PhysRevLett.96.213601}
A.~Ourjoumtsev, R.~Tualle-Brouri, and P.~Grangier, \enquote{Quantum homodyne tomography of a two-photon fock state,} {\protect\JournalTitle{Phys. Rev. Lett.}} \textbf{96}, 213601 (2006).

\bibitem{Cooper:13}
M.~Cooper, L.~J. Wright, C.~S\"{o}ller, and B.~J. Smith, \enquote{Experimental generation of multi-photon fock states,} {\protect\JournalTitle{Opt. Express}} \textbf{21}, 5309--5317 (2013).

\bibitem{Bouillard:19}
M.~Bouillard, G.~Boucher, J.~F. Ortas, \emph{et~al.}, \enquote{High production rate of single-photon and two-photon fock states for quantum state engineering,} {\protect\JournalTitle{Opt. Express}} \textbf{27}, 3113--3120 (2019).

\bibitem{Kawasaki:22}
A.~Kawasaki, K.~Takase, T.~Nomura, \emph{et~al.}, \enquote{Generation of highly pure single-photon state at telecommunication wavelength,} {\protect\JournalTitle{Opt. Express}} \textbf{30}, 24831--24840 (2022).

\bibitem{Inoue-5G}
A.~Inoue, T.~Kashiwazaki, T.~Yamashima, \emph{et~al.}, \enquote{{Toward a multi-core ultra-fast optical quantum processor: 43-GHz bandwidth real-time amplitude measurement of 5-dB squeezed light using modularized optical parametric amplifier with 5G technology},} {\protect\JournalTitle{Applied Physics Letters}} \textbf{122}, 104001 (2023).

\bibitem{PhysRevResearch.5.033156}
T.~Sonoyama, K.~Takahashi, B.~Charoensombutamon, \emph{et~al.}, \enquote{Non-gaussian-state generation with time-gated photon detection,} {\protect\JournalTitle{Phys. Rev. Res.}} \textbf{5}, 033156 (2023).

\bibitem{Yukawa:13}
M.~Yukawa, K.~Miyata, T.~Mizuta, \emph{et~al.}, \enquote{Generating superposition of up-to three photons for continuous variable quantum information processing,} {\protect\JournalTitle{Opt. Express}} \textbf{21}, 5529--5535 (2013).

\bibitem{Bock:16}
M.~Bock, A.~Lenhard, C.~Chunnilall, and C.~Becher, \enquote{Highly efficient heralded single-photon source for telecom wavelengths based on a ppln waveguide,} {\protect\JournalTitle{Opt. Express}} \textbf{24}, 23992--24001 (2016).

\bibitem{PhysRevLett.109.230503}
A.~Mari and J.~Eisert, \enquote{Positive wigner functions render classical simulation of quantum computation efficient,} {\protect\JournalTitle{Phys. Rev. Lett.}} \textbf{109}, 230503 (2012).

\bibitem{PhysRevA.64.012310}
D.~Gottesman, A.~Kitaev, and J.~Preskill, \enquote{Encoding a qubit in an oscillator,} {\protect\JournalTitle{Phys. Rev. A}} \textbf{64}, 012310 (2001).

\bibitem{PhysRevLett.120.073603}
H.~Le~Jeannic, A.~Cavaill\`es, K.~Huang, \emph{et~al.}, \enquote{Slowing quantum decoherence by squeezing in phase space,} {\protect\JournalTitle{Phys. Rev. Lett.}} \textbf{120}, 073603 (2018).

\bibitem{Endo:23}
M.~Endo, R.~He, T.~Sonoyama, \emph{et~al.}, \enquote{Non-gaussian quantum state generation by multi-photon subtraction at the telecommunication wavelength,} {\protect\JournalTitle{Opt. Express}} \textbf{31}, 12865--12879 (2023).

\bibitem{Cahall:17}
C.~Cahall, K.~L. Nicolich, N.~T. Islam, \emph{et~al.}, \enquote{Multi-photon detection using a conventional superconducting nanowire single-photon detector,} {\protect\JournalTitle{Optica}} \textbf{4}, 1534--1535 (2017).

\bibitem{Endo:21}
M.~Endo, T.~Sonoyama, M.~Matsuyama, \emph{et~al.}, \enquote{Quantum detector tomography of a superconducting nanostrip photon-number-resolving detector,} {\protect\JournalTitle{Opt. Express}} \textbf{29}, 11728--11738 (2021).

\bibitem{Provaznik:20}
J.~Provazn\'{i}k, L.~Lachman, R.~Filip, and P.~Marek, \enquote{Benchmarking photon number resolving detectors,} {\protect\JournalTitle{Opt. Express}} \textbf{28}, 14839--14849 (2020).

\bibitem{TES-ns}
A.~Lamas-Linares, B.~Calkins, N.~A. Tomlin, \emph{et~al.}, \enquote{{Nanosecond-scale timing jitter for single photon detection in transition edge sensors},} {\protect\JournalTitle{Applied Physics Letters}} \textbf{102}, 231117 (2013).

\bibitem{SNSPD}
G.~N. Gol’tsman, O.~Okunev, G.~Chulkova, \emph{et~al.}, \enquote{{Picosecond superconducting single-photon optical detector},} {\protect\JournalTitle{Applied Physics Letters}} \textbf{79}, 705--707 (2001).

\bibitem{kawasaki2024highrate}
A.~Kawasaki, R.~Ide, H.~Brunel, \emph{et~al.}, \enquote{High-rate generation and state tomography of non-gaussian quantum states for ultra-fast clock frequency quantum processors,}  (2024).

\bibitem{ILvovsky_2004}
A.~I. Lvovsky, \enquote{Iterative maximum-likelihood reconstruction in quantum homodyne tomography,} {\protect\JournalTitle{Journal of Optics B: Quantum and Semiclassical Optics}} \textbf{6}, S556 (2004).

\end{thebibliography}

\end{document}